\documentclass[aoas,MSNbibl,nameyear,dvips]{arximspdf}

\doi{10.1214/12-AOAS606} 
\volume{6}
\issue{4}
\pubyear{2012}
\firstpage{1349}
\lastpage{1351}

\begin{document}
\begin{frontmatter}

\title{Section on the special year for mathematics of planet earth (MPE 2013)}
\runtitle{Mathematics of planet earth 2013}

\begin{aug}
\author{\fnms{Tilmann} \snm{Gneiting}\corref{}\ead[label=e1]{t.gneiting@uni-heidelberg.de}\ead[label=u1,url]{http://www.foo.com}}
\runauthor{T. Gneiting}
\affiliation{Universit\"at Heidelberg}
\address{Institut f\"ur Angewandte Mathematik\\
Universit\"at Heidelberg \\
Im Neuenheimer Feld 294 \\
69120 Heidelberg \\
Germany \\
\printead{e1}\\
\printead{u1}}
\end{aug}

\received{\smonth{9} \syear{2012}}


%

\end{frontmatter}

Dozens of research centers, foundations, international organizations
and scientific societies, including the Institute of Mathematical
Statistics, have joined forces to celebrate 2013 as a special year for
the Mathematics of Planet Earth. In its five-year history, the
\textit{Annals of Applied Statistics} has been publishing cutting edge
research in this area, including geophysical, biological and
socio-economic aspects of planet Earth, with the special section on
statistics in the atmospheric sciences edited by \citet{FueGutSte08}
and the discussion paper by \citet{McSWyn11} on
paleoclimate reconstructions [\citet{Ste11}] having been highlights.

As a prelude to the special year for the Mathematics of Planet Earth,
and welcoming the concurrent International Year of Statistics, the
December 2012 issue of the \textit{Annals of Applied Statistics}
features a special section dedicated to statistical aspects of the
study of planet Earth. The section is comprised of ten papers that
span the four themes of the special year, \textit{A~Planet to Discover},
\textit{A~Planet Supporting Life}, \textit{A~Planet Organized by Humans}
and \textit{A~Planet at Risk}.

Three of the papers in this section relate to the history of planet
Earth. Reitan, Schweder and Hendriks (\citeyear{ReiSchHen12}) look into
the deep past,
studying time series of cell size evolution in marine algae,
Er\"ast\"o et al. (\citeyear{Eraetal12}) merge distinct paleoclimate
reconstructions,
and Baggaley et al. (\citeyear{Bagetal12}) consider population dynamics
in the late
Stone Age. \citet{ReiSha12}, Sigrist, K\"unsch and Stahel
(\citeyear{SigKunSta12}), \citet{CooDavNav12} and \citet{JonGelJon12}
study the atmosphere and the oceans of our planet,
looking at output from regional climate models, short term predictions
of precipitation, air pollutants and wave direction data,
respectively. Biological aspects of planet Earth are addressed by
\citet{IllSrbRue12} who consider rainforest ecosystems
and the foraging behavior of a particularly popular inhabitant of our
planet, the koala. Finally, \citet{Chi12} and \citet
{Lahetal12} set
out to solve problems of prediction and estimation, respectively, that
arise in transportation engineering.

The challenges posed by a planet at risk have been a major driver in
the development of statistical theory and methodology, and the papers
in this special section document the use of state of the art
techniques in addressing critical real world problems. Not
surprisingly, time series analysis, spatial and spatio-temporal
statistics play prominent roles in the special section papers,
including the use of point processes, Gaussian random fields,
max-stable processes and stochastic differential equations, with
functional data methods frequently addressing similar tasks. One
current challenge is to develop parsimonious, physically realistic
models for multivariate global data that can handle complex
nonstationarities in space and time [\citet{JunSte08}, \citet
{BolLin11}].
\citet{ReiSha12} and Cooley, Davis and Naveau (\citeyear{CooDavNav12})
develop the fast-paced field of extreme value statistics, including
its fruitful and important links to spatial statistics, which have
recently been reviewed by \citet{DavPadRib12}. Another
major methodological challenge lies in the marriage of analytic,
numerical and statistical techniques in inference or forecasting
problems, as discussed or touched upon by Baggaley et al. (\citeyear
{Bagetal12}),
\citet{ReiSha12}, Jona-Lasinio, Gelfand and Jona-Lasinio (\citeyear
{JonGelJon12}) and
Sigrist, K\"unsch and Stahel (\citeyear{SigKunSta12}).

State of the art applied statistical work is inevitably computational.
Developments of note in this context include the advent of the
integrated nested Laplace approximation [INLA; \citet{RueMarCho09}]
technique in Bayesian computing, which \citet{IllSrbRue12} make
available for fitting complex spatial point processes, and the
ubiquity of massive data sets, which require the adaptation of
classical techniques, as explored by \citet{Lahetal12} in the case
of bootstrap resampling.

Undoubtedly, mathematical and statistical techniques play key roles in
the multidisciplinary scholarly efforts that address the challenges
faced by planet Earth. At the \textit{Annals of Applied Statistics}, it
is our continued goal to publish cutting edge research that can help
resolve these critical issues. Simultaneously, our sister journal
\textit{Statistical Science} is preparing a special issue with invited
reviews of probabilistic and statistical facets of the scientific
study of our planet.


%

\printaddresses

\end{document}